# Secure Wireless Internet of Things Communication using Virtual Private Networks

Ishaan Lodha[1], Lakshana Kolur[2], K Sree Hari[3], and Prasad Honnavalli[4]

[1]PES University, Outer Ring Road, Bangalore.
`¹ishaanlodha@gmail.com`

[2]PES University, Outer Ring Road, Bangalore.
`²kolurlakshana@gmail.com`

[3]PES University, Outer Ring Road, Bangalore.
`³sreehari98@gmail.com`

[4]PES University, Outer Ring Road, Bangalore.
`⁴prasad.honnavalli@gmail.com`

**Abstract.** The Internet of Things (IoT) is an exploding market as well as a important focus area for research. Security is a major issue for IoT products and solutions, with several massive problems that are still commonplace in the field. In this paper, we have successfully minimized the risk of data eavesdropping and tampering over the network by securing these communications using the concept of tunneling. We have implemented this by connecting a router to the internet via a Virtual Private network while using PPTP and L2TP as the underlying protocols for the VPN and exploring their cost benefits, compatibility and most importantly, their feasibility. The main purpose of our paper is to try to secure IoT networks without adversely affecting the selling point of IoT.

## 1   Introduction

IoT is the connection of everyday mundane devices and things to a network like the internet. IoT has gained immense popularity in the recent years due to the low cost of and easily deployed sensors and actuators which can easily be controlled by micro controllers and availability of IPv6 addresses.

IoT is also one of the largest source of data in the world. It is estimated to have generated more than 500 ZB by the end of 2019[1]. The data will be mainly sensor data from the manufacturing and health care industries as they account for over 70% of IoT devices. Only 0.06% of total devices that can be connected to the internet are actually connected, which shows the potential for a huge IoT data

## 2      USP of IoT - Challenges for Security

The main selling points for the widespread and easy adoption of IoT are the very reasons that hinder its security and don't allow easy implementation of conventional security procedures on them. Thus any security procedure that is designed for IoT devices should be in tandem with these USPs and complement them in order for these procedures to be adopted by the industry.

- Small Size: IoT components are typically very small with minimal processing power. They have the minimum processing capability to serve their purpose, which allows them to remain cheap and small. They are not capable of executing conventional forms of heavy encryption and other security procedures.
- Energy Efficiency: IoT devices are predominantly battery powered and consume minuscule amounts of power. In adding security features we cannot drastically increase the power consumption of the devices as it would render them incompetent in the market [2]. The small battery also accounts for its small size and low cost which enable their large scale deployment.
- Usage Lifetime and Accessibility: IoT devices are deployed for considerably long periods of time of up to 5 years without almost any direct user contact or maintenance intervention in this period. Thus the methods implemented should not require users to access the IoT device for anything in this duration. They are also deployed in remote inaccessible locations, need for repeated intervention would be a hindrance to this cause.
- Cost: IoT devices are generally cheap and thus if implementation of security solutions makes them considerably more expensive then it would not perform well in the market, thus the cost effect of security features has to be minimal.

Thus several constraints are in place in making IoT networks secure without disrupting the present status quo in the industry. It has been attempted to secure these networks with minimum impact on the status quo and thus device a solution acceptable to and implementable by the industry.

## 3      Approaches and their Feasibility

The most common approaches to IoT security that are being extensively explored in ongoing research are as follows:

### 3.1      Hardware Implemented Security

Implement encryption on the silicon chip so that end-to-end communication is secure from the beginning. These security chips essentially give you a trusted environment that can be used for what is called a 'hardware root of trust'. These have a very specialized operating system, a specialized environment, that is built into that chip, all designed from scratch with security as a top priority. The first layer of security is on-device mitigation and hardware-implemented security is one way of achieving that. It is for security over the application layer and despite the network being secure if the end devices are not, attacks and threats are possible.

### 3.2 Device Authentication

There is a conflict in the implementation of device authentication, that is between hoe secure the authentication technique is, against how practical it is to implement it onto a small, low-cost IoT device. We need to apply the Principle of Adequate Protection where we need to choose a method that is a balance of both, that is secure enough but also cost-effective. The RSA key used for communication like ssh (secure shell) between hosts can be used for authentication of IoT devices and the recipient host. The drawback of this method is that the RSA key is large in size and its generation is computationally intensive because it has to generate a large number of bits. This is not feasible to be implemented on minute IoT devices since they need to be energy effective. [3]

### 3.3 Light Weight Encryption

There are various requirements to decide what sort of cipher to use: which are size, power consumption and processing speed (throughput, delay). Once we judge the edge device based on these parameters, we decide on using either a symmetric or an asymmetric encryption algorithm. [4] Symmetric encryption is where only one key is used to encrypt as well as decrypt information and asymmetric encryption is where there are two cryptographic keys, namely, a public key as well as a978 private key where the public key can be used to encrypt information and data but only the private key of the user can be used to decrypt it. [5]

### 3.4 Securing the Network

A plethora of companies use this process in allowing employees remote access to their business systems over the public internet. Virtual Private Networks give a virtual direct connection between systems over the public internet. They extend personal networks over the internet. By connecting the IoT devices to the edge devices over a VPN we can ensure that the data is not eavesdropped on or altered en route. This is a tried and tested method of transferring sensitive data over an unsecured network and does not add computational or temporal overhead.

# 4 Tunneling

The mechanism of the tunneling protocol is as follows: it takes the part of the packet that contains information and uses it to store the packets that actually assist in solving the problem. The protocols used by tunneling are layered, which consist of protocols such as OSI or TCP/IP. The packets that assist in service are known as payloads, these operate at a layer which is one below that of the parent packets in which they are encapsulated. [6] Proposed work uses tunneling protocol as a way to create a virtual private network. It hides all communication between the hosts connected to it by way of encryption and hiding the hosts, thus creating an illusion of the absence of any devices or communication.
The most commonly used protocols for tunneling are:

## 4.1 Point to Point Tunneling Protocol

Point to point tunneling protocol also known as PPTP. The main concept of this protocol is a client-server design. This runs on the data-link layer in the OSI hierarchy. This protocol is widely used to tunnel data over the internet. In order to initiate the connection, the user launches a point to point tunneling protocol client that in turn establishes a link to their internet provider. After this, there is a TCP connection established between the client and the server of the Virtual Private Network. It makes use of port 1723 and the concept of GRE (General Routing Encapsulation) to create the tunnel. This setup can also be done over a local network. The next step is the information flow. [7]

## 4.2 Layer 2 Tunneling Protocol

The objective of this protocol is to use an encryption mechanism in conjunction with it that is passed through the tunnel in order to achieve security of data. This is also an improvement over the Point to Point Tunneling Protocol. The protocol it uses to send the data and the header of L2TP is UDP. The reason for choosing UDP over TCP is so that we can avoid the "TCP meltdown problem". The point to point sessions are established within an L2TP tunnel. The protocol doesn't provide any sort of confidentiality by itself. Protocols such as IPsec are used to secure the L2TP tunnel. Together they are known as L2TP/IPsec. There are two ends of any L2TP tunnel. They're known as the LNS and the LAC. The LNS or the L2TP Network Server keeps searching for new tunnels. Once a connection is setup the transfer of data is full duplex. There are upper level protocols which are used in order to use this protocol for networking, this is done using an L2TP session. Either the L2TP network server or the L2TP Access Concentrator can send a request for a session. Since each session is independent from the other, we can set up a large number of virtual networks through one tunnel. There may be a constraint since each packet data should not be greater than the maximum transmission unit of that datagram. The types of packets are similar to the ones in PPTP, control and data. Here there is a guarantee that the protocol gives when transmitting control packets. If we need reliability for data packets, it must be done by the protocols that run inside L2TP.

## 5     Campus Building Network

In an attempt to implement our VPN based security concepts, we tried to tackle another real life problem around us, that of physical security of the University campus. The system we have implemented is a secure IoT system with wireless sensor networks which serve as a tool for the estate management for remotely checking and controlling infrastructure. For purpose of this tool, we have divided the rooms of the campus into two categories, one comprising classrooms, seminar halls, washrooms and so on which are considered non sensitive and should be accessible to all students while the second category is the one with sensitive rooms and to which only select persons should have access which include the likes of offices of faculty members, store rooms, examination offices etc. For the scope of this paper, we will concentrate on non sensitive type of rooms only. With respect to these rooms, the tool should be able to detect occupancy and at the end of the college day should lock a vacant room and notify the estate management about occupied rooms post college hours. It should also automatically switch off all electrical appliances of unoccupied classrooms. The estate management should also be able to give overriding commands for each and every room from a central management system. The communication between sensors/actuators and the decision making server must be wireless and secure.

The IoT network comprised of NodeMCUs with on chip esp8266 Wi-Fi module and Arduino UNO with external esp8266 Wi-Fi module as the micro controllers. We have used a variety of micro controllers to demonstrate scalability in that respect. Each room has a set of sensors comprising humidity sensor, passive IR motion sensor and temperature sensor. These sensors combined together give a stock of the situation in the classroom and about its occupancy. Each room also has a set of actuators, which include a door lock (electromagnetic or servo based), and AC relays to control the electrical appliances. The sensors log their reading in the server every second along with the state of the actuators, i.e., if the room is locked or not and if the appliances are on or off. The server reports the state of the rooms to user and saves the incoming data in a log file along with a room ID for each room. According to some rules the server sends back instructions to each room's micro controllers such as to switch on the appliances or to lock the room. The micro controllers are programmed to automatically switch off the appliances if the room is empty and if the server sends a command to lock the room but if the room is occupied then the micro controllers will not lock that room but send an error back to the server for the security to go check the room.

The architecture of this system has all sensors and actuators connected to micro controllers and the Wi-Fi module of both micro controllers connected a router. Multiple adjacent rooms share a common secure router and this router is connected to the public internet and over a VPN on the internet to the central server. For the sake of the prototype, the secure router was connected to another unsecured open router which mimicked the public internet. We tried both Point to Point Tunneling Protocol (PPTP) and Layer 2 Tunneling Protocol (L2TP) as the underlying protocol of the VPN. For PPTP we used the pated tool on Ubuntu machines and for L2TP we user the MacOS Server application which created a L2TP VPN server. The secure router connects to the VPN server and allows secure tunneled communication with the server. We connected another computer to the unsecured router or the internet and used Wireshark to sniff for packets being transmitted through that router. In both tunneling protocols we could not sniff any packets while without the VPN, all packets could be sniffed, which meant we could intercept them and even use ARP spoofing to even mount a MITM attack. Thus the use of tunneling returned positive results.

The server is written in Java and uses swing GUI toolkit and several threads. Each secure router has port forwarding enabled to allow each esp8266 module to have full duplex communication with the central server. The central server has a Hash Table of room IDs mapped to the IP address of the secure router for that room and port numbers for the port forwarding from the router to that room's micro controllers. The GUI has a text box for room ID and buttons to send commands like Lock/Unlock the room and turn on/off the appliances and it returns the status of execution of these commands and errors faced if any. It also has an option to get the status of a room which is retrieved from the most recent log and printed in proper format, giving status of the lock, appliances, occupancy and temperature. The server has also been configured to automatically sound the fire alarms if the temperature of the room exceeds a certain threshold. The GUI also allows the user to sound a buzzer in the entered room ID which can be used as a signal to empty the room. The format of the log is the room number hyphen is flag 0 or 1 for motion sensed or not, followed by flag for if appliances are on or not, then it has the angle of the servo which gives the status of the door lock, ending with the temperature. IT further has an administrator panel to add, change and delete room IDs and the corresponding IP address and port numbers.

## 6 Major IoT Cybersecurity Concerns

### 6.1 Network Attacks

- MITM: In network security, a man-in-the-middle attack (MITM) is an unauthorized attack where the attacker tries to intercept communication between two parties and gain access to their sensitive information. Active eavesdropping is a common type where an attacker intercepts the incoming messages and sends new messages of his own to the receiver. These types of attacks are more prevalent in IoT devices due to the lack of attention given to changing passwords, and also because they do not have any check on security certificates' validity.
- Botnets: are a myriad of devices connected by the Internet, handled by the attacker who then manipulates these devices, sometimes without their knowledge, to deploy DDOS attacks. IoT devices are more susceptible because they lack an antivirus protection layer which would probably detect the malware, that is in contrast, not detected when dormant, hence greatly easing the process of deploying DDOS attacks because of the reduced security complexity aspect. [8]

## 6.2    Other Attacks

All the software attacks like viruses, worms, trojans etc., that can be used on a computer system can also be used to attack IoT devices. In IoT devices this threat is amplified due to their minimalistic nature and size. They do not have any protection in form of malware and cannot support them as well. IoT devices can also be subjected to side channel attacks where the attacker uses data like computational power used in encryption to statistically guess the algorithm and key used. [9]

## 7    Future Work

A defense in depth approach with multiple security mechanisms in place would be an ideal setup. Light weight encryption of data before it is sent to the router to be transmitted via the tunnel covers all bases and makes the network completely secure. Light weight encryption is also the most suitable combination as multitude of research has been undertaken in the same and is comparatively more feasible. It could also be device authentication on connecting to the VPN and before communication can commence via the tunnel.

## 8    Conclusion

Wireless networks of IoT devices have been secured by hiding the devices on a virtual private network. The proposed work connects IoT devices to the main servers through a VPN server which hides the communication between them from the public internet by way of VPN's underlying encryption. L2TP and PPTP have been used, tested and discussed for the given scenario. The proposed security works on the principle of concealment. It has a classroom network to control locks and appliances which are controlled by a central admin server, the communication between them is over an open wifi, to mimic the public internet, and has been secured by use of a VPN server that facilitates the communication.

**References**

1. Cisco Global Cloud Index: Forecast and Methodology, 2016–2021 White Paper, November 2018.
2. Asma Haroon et al, Constraints in the IoT: The World in 2020 and Beyond, 2016, pp.1--2.
3. Zhang, Jiansong, et al. "Proximity based IoT device authentication." IEEE INFOCOM 2017-IEEE Conference on Computer Communications. IEEE, 2017.
4. Sridhar S, Smys S. Intelligent security framework for iot devices cryptography based end-to-end security architecture. In 2017 International Conference on Inventive Systems and Control (ICISC) 2017 Jan 19 (pp. 1-5). IEEE.
5. M. Katagi and S. Moriai, "Lightweight cryptography for the internet of things," Sony Corporation, pp. 2–4, 2008.
6. Riahi Sfar, Arbia \& Challal, Yacine \& Natalizio, Enrico \& Chtourou, Zied \& Bouabdallah, Abdelmadjid. (2013). A Systemic Approach for IoT Security. Proceedings - IEEE International Conference on Distributed Computing in Sensor Systems, DCoSS 2013. 351-355. 10.1109/DCOSS.2013.78.
7. Miklós, György, and Janos Harmatos. "Mobile layer 2 virtual private network over internet protocol networks." U.S. Patent No. 9,173,153. 27 Oct. 2015.
8. Nawir, Mukrimah, et al. "Internet of Things (IoT): Taxonomy of security attacks." 2016 3rd International Conference on Electronic Design (ICED). IEEE, 2016.
9. Deogirikar, Jyoti, and Amarsinh Vidhate. "Security attacks in IoT: A survey." 2017 International Conference on I-SMAC (IoT in Social, Mobile, Analytics and Cloud)(I-SMAC). IEEE, 2017.